\documentclass[pre,aps,10pt,twocolumn]{revtex4-1}

\usepackage{graphicx}

\usepackage{amsmath}
\usepackage{amsfonts}
\usepackage{amssymb}
\usepackage{amsthm}
\usepackage{mathrsfs}
\usepackage{dsfont}

\DeclareMathAlphabet{\mathpzc}{OT1}{pzc}{m}{it}

	\newcommand{\Inc}[1]{\Delta #1}

	\newcommand{\bb}[1]{\mathbb{#1}}		
	\newcommand{\mr}[1]{\mathrm{#1}}			

	\newcommand{\br}[1]{\left( #1 \right)}
	\newcommand{\brr}[1]{\left[ #1 \right]}
	\newcommand{\brrr}[1]{\left\{ #1 \right\}}
	\newcommand{\of}[1]{\!\br{#1}}
	\newcommand{\off}[1]{\!\brr{#1}}
	\newcommand{\offf}[1]{\!\brrr{#1}}
	\newcommand{\sbr}[1]{( #1 )}
	\newcommand{\sbrr}[1]{[ #1 ]}
	\newcommand{\sbrrr}[1]{\{ #1 \}}
	\newcommand{\sof}[1]{\!\sbr{#1}}
	\newcommand{\soff}[1]{\!\sbrr{#1}}
	\newcommand{\sofff}[1]{\!\sbrrr{#1}}

	\newcommand{\Sum}[2]{\sum\limits_{#1}^{#2}}

	\newcommand{\Int}[3]{\int\limits_{#1}^{#2}\mr{d}#3\,}

	\newcommand{\CondProb}[2]{\bb{P}\offf{#1 \bigg| #2}}
	\newcommand{\EA}[1]{\xpc{#1}}
	\newcommand{\Var}[1]{\mr{Var}\off{#1}}
	
	\newcommand{\TA}[2]{\overline{#1}_{#2}}
	\newcommand{\xpc}[1]{\left\langle #1 \right\rangle}
	
	\newcommand{\sProb}[1]{\bb{P}\sofff{#1}}
	
	\newcommand{\sEA}[1]{\sxpc{#1}}

	\newcommand{\sxpc}[1]{\langle #1 \rangle}
	
	\newcommand{\CondEA}[2]{\bb{E}\off{#1 \bigg| #2 }}
	\newcommand{\sCondEA}[2]{\bb{E}\soff{#1 | #2 }}

	\newcommand{\Landau}[1]{\mathpzc{O}\of{#1}}
	\newcommand{\sLandau}[1]{\mathpzc{O}\sof{#1}}
	\newcommand{\landau}[1]{\mathpzc{o}\of{#1}}
	\newcommand{\slandau}[1]{\mathpzc{o}\sof{#1}}

		\newcommand{\Min}[2]{\min\of{#1,#2}}
		\newcommand{\Max}[2]{\max\of{#1,#2}}
		\newcommand{\Abs}[1]{\left\vert #1 \right\vert}
		\newcommand{\sAbs}[1]{\vert #1 \vert}

		\newcommand{\Id}{\mathds{1}}
		\newcommand{\Ind}[2]{\Id_{#1}\of{#2}}

\newcommand{\TASD}{\TA{\Inc{X^2}\of{\tau}}{T}}

\begin{document}
	\title{Time averages in continuous time random walks}
	\date{\today}

	\author{Felix Thiel}
	\email{thiel@posteo.de}
	\affiliation{Institut f\"ur Physik, Humboldt Universit\"at zu Berlin, Newtonstra\ss e 15, 12489 Berlin, Germany}

	\author{Igor M. Sokolov}
	\email{igor.sokolov@physik.hu-berlin.de}
	\affiliation{Institut f\"ur Physik, Humboldt Universit\"at zu Berlin, Newtonstra\ss e 15, 12489 Berlin, Germany}

	\begin{abstract}
		We investigate the time averaged squared displacement (TASD) of continuous time random walks with respect to the number of steps $N$, which the random walker performed during the data acquisition time $T$.
		We prove that in each realization the TASD grows asymptotically linear in the lag time $\tau$ and in $N$, provided the steps can not accumulate in small intervals.
		Consequently, the fluctuations of the latter are dominated by the fluctuations of $N$, and fluctuations of the walker's thermal history are irrelevant.
		Furthermore, we show that the relative scatter decays as $1/\sqrt{N}$, which suppresses all non-linear features in a plot of the TASD against the lag time.
		Parts of our arguments also hold for continuous time random walks with correlated steps or with correlated waiting times.
	\end{abstract}

	\maketitle

	\section{Introduction}
		The continuous-time random walk (CTRW) with power-law distributed waiting times advanced from a specific model for charge transport in disordered semiconductors \cite{Montroll1965,Scher1975}, to one of the standard models for anomalous diffusion in general \cite{Klafter2011}.
		Just recently, CTRW has been employed to explain the anomalous behavior of a probe in granular media, \cite{Scalliet2015}.
		Contrary to a common random walker or a particle moving according to Brownian motion, a continuous-time random walker has to remain for a random time at its place before it is allowed jumping again.
		Its trajectory is piece-wise constant.
		The waiting times are independent identically distributed (iid) random variables.
		When the probability density function (PDF) of the waiting times $\psi\sof{t}$ possesses a diverging first moment, e.g. when it behaves like $\psi\of{t} \propto t^{-1-\alpha}$ with $\alpha < 1$ for large $t$, the mean squared displacement (MSD) $\sxpc{X^2\of{t}}$ of the CTRW grows anomalously, i.e. non-linearly in time.
		In this case it grows as $t^\alpha$; a case of so-called sub-diffusion, which is also observable in interacting particles on comb-like structures, \cite{Benichou2015} or in the diffusion in biological cells, \cite{Jeon2011,Weigel2011,Tabei2013}.
		On the other hand, the behavior of the time-averaged squared displacement (TASD) for a single trajectory of the process 
		\begin{equation}
				\TASD
			:=
				\frac{1}{T-\tau} \Int{0}{T-\tau}{t} \brr{ X\of{t+\tau} - X\of{t} }^2
			\label{eq:DefTASD}
		\end{equation}
		and of its ensemble-averaged analogue is linear in the lag time $\tau$, \cite{Lubelski2008,He2008}.
		The time averaging erases the anomaly of diffusion, but introduces the anomalous dependence on the total data acquisition time $T$ which is absent in the case of normal diffusion for $T$ long enough.
		Testing the behavior of different single-trajectory and ensemble properties of the displacement may be used as a tool for accepting or rejecting the CTRW as a candidate process for the behavior observed in experiment or in simulation \cite{Meroz2010}. 
		Note that such a test was applied e.g. in Refs. \cite{Neusius2008} before the full theory underlying such tests was built. 
		The linear $\tau$-behavior in the double average is present in processes with uncorrelated increments, as was shown in Ref. \cite{Massignan2014}.
		Those are the processes without ``structural disorder'', \cite{Thiel2013}.
		When they are prepared in equilibrium, the anomaly can not be observed anymore.
		
\begin{figure*}
			\includegraphics[width=0.49\textwidth]{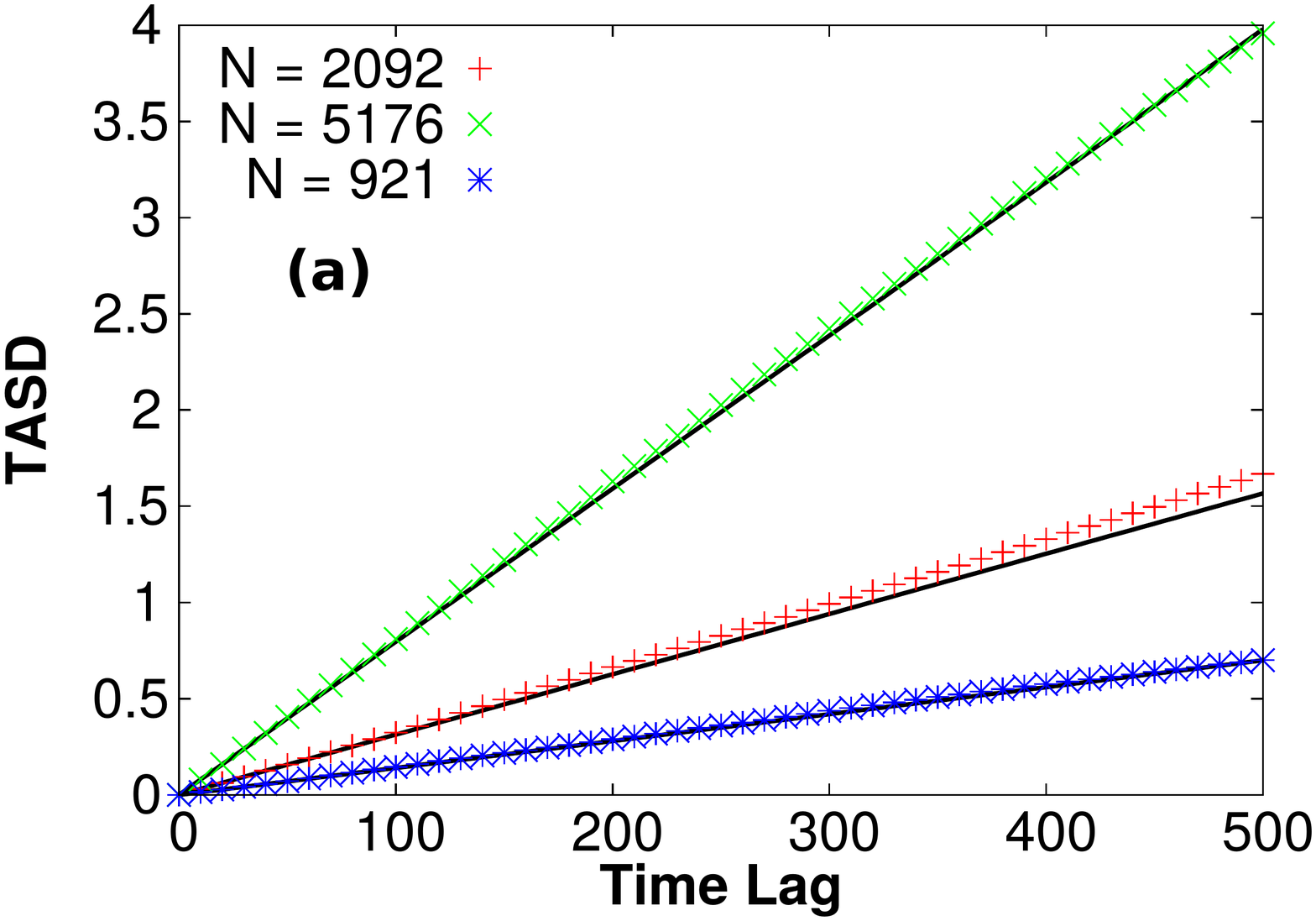} 
			\includegraphics[width=0.49\textwidth]{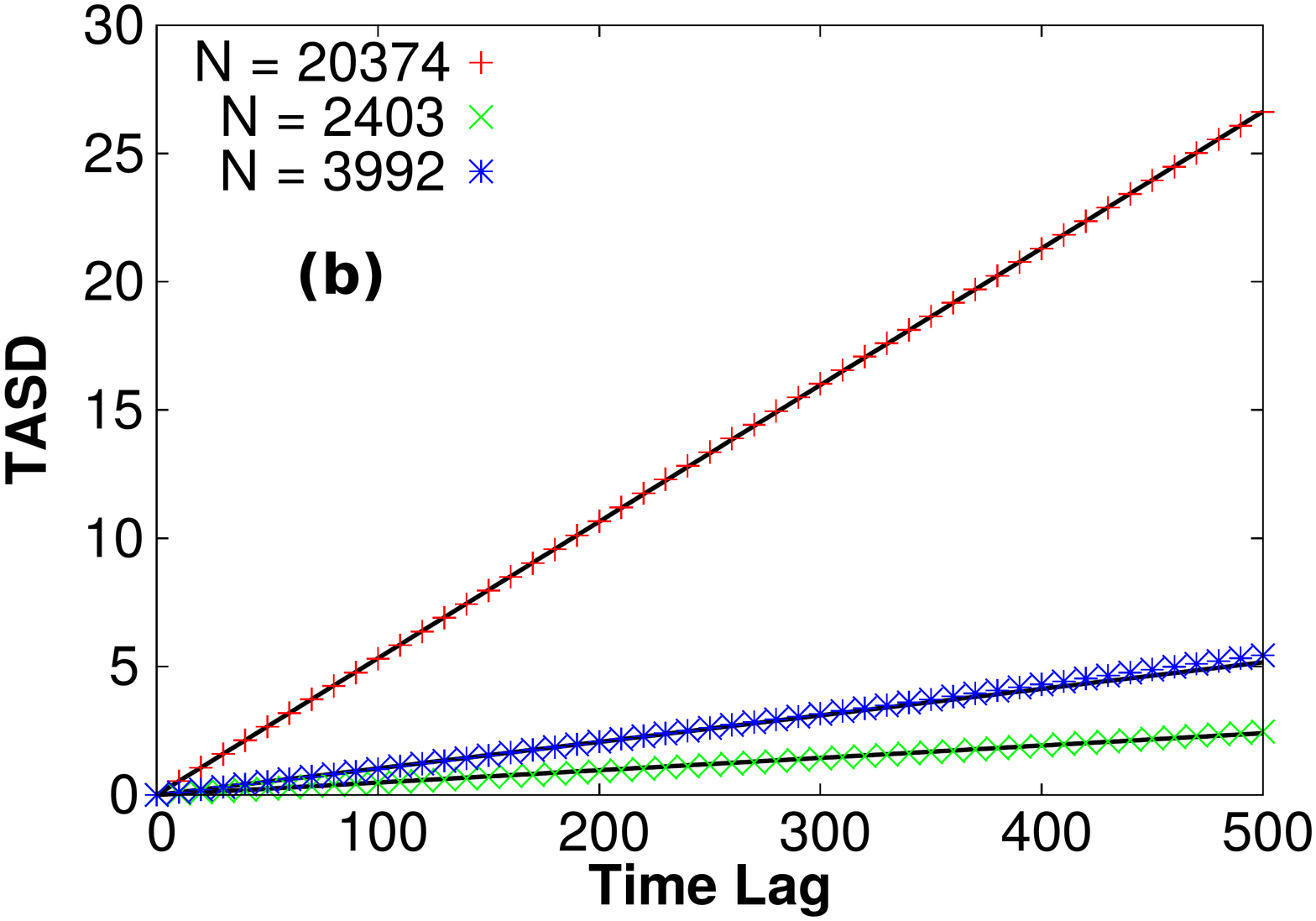} 
			\caption{
				TASD in aging CTRW and in the trap model.
				The TASD from single trajectories is plotted against the lag time $\tau$ for aging CTRW (a, left) and for the random trap model (b, right).
				Observe the almost linear behavior.
				The numerical results (symbols) are compared with the prediction from Eq.\eqref{eq:TASDL2Eq}, (black lines).
				To do this we computed the sum of the squared displacements of each trajectory.
				The lines are no fits!
				For each trajectory we provide the number of jumps $N$ in the measurement interval, which dominate the slope of the line.
				Total measurement time was $2\cdot 10^6$.
				The CTRW's the waiting times were Pareto distributed with exponent $\alpha=0.75$.
				Its aging time was $2\cdot10^5$.
				For the trap model a landscape with $4096$ traps was prepared, the trapping times were Pareto distributed with exponent $\alpha' = 0.75$.
				\label{fig:Plot}
			}
		\end{figure*}

		The value of $\TASD$, as well of other time averaged quantities like occupation times, fluctuates strongly between different realizations of the process, indicating weak ergodicity breaking \cite{Bel2005,He2008,Metzler2014-1}.
		In a \textit{given realization} of a CTRW however $\TASD$ shows an astonishingly linear dependence on $\tau$, see Fig.~\ref{fig:Plot} or Fig.~2 of Ref.~\cite{Lubelski2008}:
		Its deviations from strictly linear behavior are hardly visible by the naked eye. 
		This allows for definition of the apparent diffusion coefficients $K$, defined via 
		\begin{equation*}
				\TASD
			=
				2 K\of{T} \tau
			,
		\end{equation*}
		that fluctuates strongly from one realization to another \cite{He2008,Miyaguchi2011-1,Miyaguchi2015}.
		Its distribution exhibits a considerable amount of universality \cite{Sokolov2008}.
		It has been subject of several studies, also in other processes than CTRW, \cite{Akimoto2013,Akimoto2014}.
	
		The distribution of $K\of{T}$ was calculated in \cite{He2008}, where it was assumed that the value of $K$ is given by $K \approx \lambda^2 N\of{T} / T$, where $N\of{t}$ is the total number of steps performed up to time $t$.
		This implies that fluctuations in $K$ are dominated by those in $N$, and not by the fluctuations of the thermal histories (i.e. by the different directions of steps).
		In other words, the TASD $\TASD$ \textit{conditioned on the total number of steps} done during the data acquisition time practically does not fluctuate. 
		This statement is based on strong numerical evidence and even holds in the presence of aging, i.e. when the process was not prepared at an renewal epoch, see \cite{He2008,Schulz2014}.

		Although the linear behavior of the double average is reasonably well understood, the linearity of {\em single realization} time averages is not.
		However, a thorough theoretical understanding is imperative, as single particle tracking became a standard experimental tool.
		We will provide this in the present paper.

		Before going to the technical details of calculations which are given below let us discuss the situation in plain words using a simple picture in which the step size $\xi_n$ is $\pm a$, and assume for simplicity $\tau \ll T$.
		As the measurement time $T$ grows, the typical value of $N \propto T^{\alpha}$ grows as well, but the rate or density of steps, $N/T$, declines.
		Whence, a small interval of length $\tau$ contains at most one jump, and the displacement in this interval is either zero or $a^2$.
		The TASD in this case is proportional to the mean number of $\tau$-intervals which contain a step of which there are $N$.
		Hence, $\TASD = a^2 N \tau / (T-\tau) \approx \tau a^2 N/T$ as assumed in \cite{He2008}. 
		Conditioned on $N$, it does not fluctuate at all. 
		As we proceed to show, the actual situation is similar:
		An interval of length $\tau$ may contain much more than just a single jump, but still not a significant number of jumps.
		The linearity in $\tau$ is actually a consequence of the resting periods between the steps.

	\section{A representation of piecewise constant random walks}
		The CTRW is a process subordinated to a simple random walk under the operational time $U\of{t}$, that measures the number of steps until time $t$. 
		This operational time defines the internal clock of the walker. 
		For CTRW, this operational time is an renewal process and grows on the average as $\sxpc{U\of{t}} \sim t^\alpha$, where $\alpha$ is the exponent of the waiting time PDF.
		The MSD on the other hand grows linearly with the number of steps, so that we have $\sxpc{X^2\of{t}} \sim \sxpc{U\of{t}} \sim t^\alpha$.

		In our further discussion we use the approach based on the averaging of combinations of indicator functions which considerably simplifies the bookkeeping. 
		A similar approach was used in Refs. \cite{Miyaguchi2011-1,Miyaguchi2013,Akimoto2013,Akimoto2014}.
		An indicator function $\Ind{A}{x}$ evaluates to unity if $x\in A$ and is zero otherwise. 
		The advantage of this approach is based on the fact that any power of the indicator function is equal to the function itself, and that the product of indicator functions corresponding to two intervals is an indicator function corresponding to their intersection $\Ind{A}{x} \Ind{B}{x} = \Ind{A \cap B}{x}$. 
		These properties make the formal calculations easy. 

		The displacements of a walker (jumps, or steps) take place at the time instants $T_n$, during the waiting time intervals $[T_n,T_{n+1})$ the walker is immobile. 
		The displacement in $n$-th step is $\xi_n$. 
		The $\xi_n$ are iid random variables. 
		The distribution of $\xi_n$ is assumed to be symmetric, and to possess finite fourth moment. 
		The total displacement $X\sof{t}$ of the random walker starting at the origin at time $t=0$ is then  
		\begin{equation}
				X\of{t}
			= 
				\Sum{n=1}{\infty} \xi_n \Ind{[T_n,\infty)}{t}
			,
			\label{eq:DefX}
		\end{equation}
		where the indicator function simply bounds the summation to such numbers of steps $n$ that $T_n < t$.
		In the same way, the operational time -- the number of steps performed until time $t$ -- can be written as:
		\begin{equation}
				U\of{t}
			=
				\Sum{n=1}{\infty} \Ind{[T_n,\infty)}{t}
			.
			\label{eq:DefU}
		\end{equation}
		Using Eq.~\eqref{eq:DefX} it is particularly easy to obtain the TASD.
		First we note that the increment of $X\of{t}$ over an interval of lag time $\tau$ is given by:
		\begin{equation*}
				\Inc{X}\of{t,t+\tau}
			:=
				X\of{t+\tau} - X\of{t}
			=
				\Sum{n=1}{\infty} \xi_n \Ind{[T_n-\tau,T_n)}{t}
			.
		\end{equation*}
		A similar relation holds for $\Inc{U}\of{t,t+\tau}$, when $\xi_n$ is replaced with unity.
		To get the TASD we square the expression and integrate over $t$, see Eq.~\eqref{eq:DefTASD}:
		\begin{equation}
				\TA{\Inc{X^2(\tau)}}{T}
			= 
				\Sum{m,n=1}{\infty} \xi_m \xi_n \theta_{m,n}
			.
			\label{eq:CalcTASD}
		\end{equation}
		Here $\theta_{m,n}$ denotes the remaining integral over $t$:
		\begin{equation}
				\theta_{m,n}
			:=
				\frac{1}{T-\tau} \Int{0}{T-\tau}{t}
				\Ind{[T_m-\tau,T_m)\cap[T_n-\tau,T_n)}{t} 
			.
			\label{eq:DefTheta}
		\end{equation}
		The integrals $\theta_{m,n}$ are random variables and can be considered as elements of a random matrix.  
		The properties of this matrix are crucial for the following discussion.

	\section{Properties of the $\theta$-matrix}
		$\theta$ is a real, non-negative and symmetric matrix.
		$\theta_{m,n}$ is proportional to the length of $[T_m-\tau,T_m)\cap[T_n-\tau,T_n)\cap[0,T-\tau)$.
		Hence its maximal value is $\tau / (T-\tau)$ and it vanishes when $T_n - T_m \ge \tau$.
		Let $N = U\of{T}$ be the numbers of steps until the measurement ends.
		Then $T_N$ corresponds to the time of the last jumps before $T$.
		For any $m>N$ the interval $[T_m-\tau,T_m)$ is completely outside of $[0,T-\tau)$.
		Therefore, for given $N$, the rank of the matrix $\theta$ is at most $N$, i.e. it can be represented as an $N\times N$ matrix.
		Provided $m\le N$ and $n\le N$, its entries are given by 
		\begin{equation}
				\theta_{m,n}
			=
				\left\{ \begin{array}{ll}
						\frac{\Min{T_m}{T_n}}{T-\tau} \Ind{[0,\tau)}{\sAbs{T_n -T_m}};			&	T_m,T_n \le \tau			\\
						\frac{T - \Max{T_n}{T_m}}{T-\tau} \Ind{[0,\tau)}{\sAbs{T_n -T_m}};		&	T-\tau < T_m,T_n	\\
						\frac{\tau - \Abs{T_n - T_m}}{T-\tau} \Ind{[0,\tau)}{\sAbs{T -T_n}};	&	\text{else}
				\end{array} \right.
			\label{eq:Theta}
		\end{equation}
		All other entries vanish.
		We find that $\theta$'s diagonal entries for $\tau < T_n < T-\tau$ are given by
		\begin{equation}
				\theta_{n,n} 
			= 
				\frac{\tau}{T-\tau}
			,
			\label{eq:ThetaDiag}
		\end{equation}
		and that all other entries can be bounded by:
		\begin{equation*}
				\theta_{m,n}
			\le
				\frac{\tau}{T} \Ind{[0,\tau)}{\sAbs{T_n - T_m}}
			.
		\end{equation*}
		Later, we will need the conditional expectation $\sCondEA{\theta_{m,n}^2}{N}$; using the last equation we can estimate this quantity with:
		\begin{equation}
				\CondEA{\theta_{m,n}^2}{N}
			\le 
				\br{\frac{\tau}{T}}^2 \CondProb{\sAbs{T_n - T_m} < \tau}{N}
			,
			\label{eq:ThetaEst}
		\end{equation}
		i.e. with the probability that the $n$-th and $m$-th jump are closer than $\tau$ under the condition that the random walker performs $N$ jumps during the measurement.

		Using Bayes' theorem we can compute this probability.
		In CTRW the sojourn times are independent and the expression only depends on the difference $ \eta := \sAbs{n-m}$:
		\begin{equation}
				\CondProb{\sAbs{T_{m+\eta} - T_m} < \tau}{N}
			=
				\Int{0}{\tau}{t} \frac{\chi_{N-\eta}\of{T-t} \psi_\eta\of{t}}{\chi_N\of{T}}
			.
		\label{eq:CondProbCTRW}
		\end{equation}
		Here $\psi_\eta\of{t}$ is the PDF of the sum of $\eta$ waiting times $T_{m+\eta}-T_m$, and $\chi_n\of{t}$ is the probability to have exactly $n$ jumps in the interval up to time $t$ that starts with a renewal, for details see \cite{Klafter2011}.
		If $\tau \ll T$, we can replace the difference $T-t$ with $T$ and we can pull the quotient of $\chi$'s in front of the integral.
		The remaining integral is the cumulative distribution function (CDF) of $T_\eta$ and decays very quickly with $\eta$.
		In fact, $\eta^{-1/\alpha} T_\eta$ converges in distribution to an $\alpha$-stable random variable $Z$.
		The sought after probability is related to the CDF of that random variable and we have: $\sProb{T_\eta<\tau} = \sProb{Z<\tau \eta^{-1/\alpha}} = \slandau{e^{-C\tau^{-\alpha}\eta}}$, with some positive $C$.
		The asymptotic relation is Eq. (6.2), Theorem 1, from chapter XIII.6 of \cite{Feller1971}.
		As the cumulative probability decays faster than exponentially in $\eta$, only small $\eta$ values are important.
		In this small-$\eta$ regime, $\chi_{N-\eta}\of{T} / \chi_{N}\of{T}$ is close to unity.
		Hence, we have in summary:
		\begin{equation}
				\CondEA{\theta_{m,n}^2}{N}
			=
				\br{\frac{\tau}{T}}^2
				\landau{e^{-C\tau^{-\alpha}\eta}}
			.
		\label{eq:ThetaEstCTRW}
		\end{equation}

	\section{The time averaged squared displacement}
		We have represented the time average as a double sum in Eq.~\eqref{eq:CalcTASD}.
		Fixing $N = U\of{T}$, we can split the sum into diagonal and off-diagonal terms:
		\begin{equation*}
			\TASD | N
			=
			\Sum{n=1}{N} \xi_n^2 \theta_{n,n} 
			+
			\Sum{m\ne n}{N} \xi_m \xi_n \theta_{m,n}
			.
		\end{equation*}

		The first sum obviously grows linearly with $N$.
		The diagonal elements of $\theta$ can be replaced with $\tau/(T-\tau)$.
		This is correct for all $U\of{\tau} < n < U\of{T-\tau}$, whence the error is proportional to the number of jumps in the intervals $[0,\tau)$ and $[T-\tau,T)$.
		Since this number is also controlled by the expression \eqref{eq:CondProbCTRW}, it can be neglected in comparison to the large number $N$.
		We have 
		\begin{equation*}
			\Sum{n=1}{N} \xi_n^2 \theta_{n,n}
			=
			\frac{\tau}{T-\tau} \Sum{n=1}{N} \xi_n^2
			+ \landau{N}
			.
		\end{equation*}

		What about the off-diagonal terms?
		Using Eq.~\eqref{eq:Theta}, neglecting the very first and the very last jumps, we see that this sum counts all pairs of jumps that are closer than $\tau$.
		\begin{equation*}
			\Sum{m\ne n}{N} \xi_m \xi_n \frac{\tau - \sAbs{T_n - T_m}}{T-\tau} \Ind{[0,\tau)}{\sAbs{T_n - T_m}}
			.
		\end{equation*}
		Using some estimates, we proceed to show that this sum is of order $\sLandau{\sqrt{N}}$.
		To see this, we examine its mean square at fixed $N$.
		This means we average over the $\xi$'s and use their independence:
		\begin{align}
				\xpc{\xi_k\xi_l\xi_m\xi_n}
			\nonumber = &
				\lambda^4 \left\{ 
					\kappa \delta_{k,l} \delta_{l,m} \delta_{m,n}
					+ \br{1 - \delta_{k,l} \delta_{l,m} \delta_{m,n} } \times 
				\right.
			\\ & \left. \times 
					\brr{
						\delta_{k,l} \delta_{m,n} 
						+ \delta_{k,m} \delta_{l,n} 
						+ \delta_{k,n} \delta_{l,m}
					}
				\right\}
			.
			\label{eq:XiKurt}
		\end{align}
		$\lambda := \sqrt{\sEA{\xi^2}}$ is the typical step length, and $\kappa$ is the kurtosis of $\xi$, i.e. the ratio of its fourth moment and of its second moment squared.
		Noting that no terms with $\xi_m^4$ occur in the sum, we obtain:
		\begin{align*}
			&
				\CondEA{
					\brrr{ \Sum{m\ne n}{} \xi_m \theta_{m,n} \xi_n }^2
				}{N}
			\\ =  &
				2 \lambda^4 \Sum{m \ne n}{N} \CondEA{\theta^2_{m,n}}{N}
			= 
				4 \lambda^4 \Sum{m=1}{N-1} \Sum{n=m+1}{N} \CondEA{\theta^2_{m,n}}{N}
			\\ \le &
				\frac{4 \lambda^4 \tau^2}{T^2} 
				\Sum{\eta=1}{N-1} \br{N-\eta} \CondProb{T_{m+\eta} - T_m \le \tau}{N}
			\\ \le &
				\br{ \frac{2 \lambda^2\tau}{T} }^2 N
				\Sum{\eta=1}{\infty} \landau{e^{-C\tau^{-\alpha}\eta}}
			\\ = &
				\br{ \frac{2 \lambda^2\tau}{T} }^2 \frac{N \tau^\alpha}{C}
			=
				\Landau{N}
			.
		\end{align*}
		Here, we used again $\eta = \sAbs{n-m}$ and rearranged the double sum.
		Then, Eq.~\eqref{eq:ThetaEst} and the asymptotic expression, Eq.~\eqref{eq:ThetaEstCTRW}, for the $\theta$-matrix is used.
		The difference $N-\eta$ is bounded by $N$, and the summation limit is put to infinity.
		Finally the summation is performed, which bears an additional factor in $\tau$.
		We thus have shown that the expected square of the off-diagonal terms is of order $\sLandau{N}$, whence the sum of off-diagonal terms itself is of order $\sLandau{\sqrt{N}}$.
		
		Putting together our last two arguments, we have in the mean-square sense:
		\begin{equation}
				\TASD|N
			=
				\frac{\tau}{T-\tau} 
				\Sum{n=1}{N} \xi^2_n
				+ \Landau{\sqrt{N}}
			.
		\label{eq:TASDL2Eq}
		\end{equation}
		This equation is our main result: The TASD is proportional to the sum of squared displacements.
		Each step contributes equally to the sum, namely $\xi_n^2 \tau/(T-\tau)$.
		The equation holds for {\em each} realization in the sense that any deviation becomes more and more unlikely as $N$ grows.

		As all increments possess a finite fourth moment (i.e. a finite kurtosis $\kappa$), $\xi^2$ has a finite second moment and the central limit theorem applies to their sum.
		Whence, the sum grows linear in $N$.

		This $N$-linearity is a consequence of the independence of the increments $\xi$.
		The linearity in $\tau$ on the other hand is a consequence of the process being {\em constant} between the jumps.
		Eq.~\eqref{eq:ThetaDiag} is a crucial relation that does not hold for a random walker that is mobile in between the steps, e.g. a L\'evy walker.

		Any non-linearity in $\tau$ is hidden in the off-diagonal sum, which counts the pairs of jumps that are closer than $\tau$.
		For CTRW, this sum is negligible {\em regardless of the actual number of jumps in a $\tau$-interval}.
		Hence, our ``hand-waving argument'' from the first section is even stricter than necessary.
		However, we do require that $\tau \ll T$ in the derivation of Eq.~\eqref{eq:ThetaEstCTRW}.

		If the displacements $\xi$ possess a finite fourth moment, the central limit theorem ensures that the TASD and hence the apparent diffusion constant at large but fixed $N$ is a Gaussian random variable.
		In this sense, its fluctuations are governed by the fluctuations in $N$, but not those of the thermal history, which are the fluctuations in the $\xi_n$.

		At fixed $N$, the TASD's mean is 
		\begin{equation}
			\CondEA{\TASD}{N}
			=
			\frac{\lambda^2 N}{T-\tau} \tau
			+\Landau{\sqrt{N}}
			,
		\label{eq:TASDMean}
		\end{equation}
		and its variance is
		\begin{equation}
			\Var{\TASD \big| N}
			=
			\br{\kappa - 1}
			\br{ \frac{\tau \lambda^2}{T-\tau} }^2
			N
			+ \Landau{N}
			=
			\Landau{N}
			.
		\label{eq:TASDVar}
		\end{equation}
		Hence relative fluctuations of the TASD die out as the random walker performs more and more jumps.
		In this way, deviations from the linear $\tau$-behavior in a realization of the TASD also vanish.
		This is true provided the increments' kurtosis $\kappa$ is finite.

		We stress that in estimating the off-diagonal terms, we did not refer to the time of the first jump, $T_1$, nor to its distribution.
		Therefore our arguments apply in particular to ageing CTRW.

	\section{A note on CTRW with correlated jumps}
		In our derivation, we explicitly used the independence of the displacements $\xi$.
		This strong requirement can be relaxed; we can admit correlated steps and consider a process subordinated to the correlated random walk.
		If we assume that the sequence of steps is stationary, then the sum $\xi_{n+1} + \xi_{n+2} + \hdots + \xi_{n+\eta}$ has the same distribution as $\xi_1 + \xi_2 + \hdots + \xi_{\eta}$.
		With the same operational time as before, the modified $X\of{t}$ is a process subordinated to the one with stationary increments (compare with the subordinated fractional Brownian motion of \cite{Thiel2014-1}). 
		In this case there exists a specific absolute moment of the displacement, which behaves especially simple and is additive in the number of steps: It is the fundamental moment \cite{Thiel2013} with index $\gamma_F$ such, that
		\begin{equation*}
				\EA{\Abs{ \Sum{m=1}{\eta} \xi_m }^{\gamma_F}}
			=
				\sigma^{\gamma_F} \eta
			,
		\end{equation*}
		or alternatively,
		\begin{equation*}
			\CondEA{ \Abs{\Delta X\of{t,t+\tau}}^{\gamma_F} }{U\of{t}}
			=
				\sigma^{\gamma_F} \brr{ U\of{t+\tau} - U\of{t} }
			.
		\end{equation*}
		The value of $\sigma$ is a particularly useful measure for the fluctuations in such a process.
		The case of uncorrelated steps discussed above corresponds to the special case $\gamma_F = 2$.
		The time average of the fundamental moment is dominated by the fluctuations of $U\of{T}$, as well.
		To see this, repeat the above arguments to obtain:
		\begin{align*}
			\CondEA{ \Abs{\Delta X\of{t,t+\tau}}^{\gamma_F} }{U\of{t}}
			= &
			\frac{\sigma^{\gamma_F}}{T-\tau} \Int{0}{T-\tau}{t} 
			\brr{ U\of{t+\tau} - U\of{t} }
			\\ = &
			\sigma^{\gamma_F} \Sum{m=1}{\infty} \theta_{m,m}
			.
		\end{align*}
		In the last line, Eq.~\eqref{eq:DefU} was used.
		As we have seen before, the series evaluates to $N\tau/(T-\tau)$, whence:
		\begin{equation}
			\CondEA{ \Abs{\Delta X\of{t,t+\tau}}^{\gamma_F} }{N}
			=	
			\br{ \frac{N \sigma^{\gamma_F}}{T-\tau} } \tau
			+ \landau{N}
			.
		\end{equation}
		This generalizes the treatment to correlated CTRW and shows that the time averaged fundamental moment of the process behaves linearly in the time lag.
		Information about its fluctuations can, however, not be inferred without more precise knowledge about the correlation structure of the process.

	\section{A note on CTRW with correlated waiting times}
		The off-diagonal sum can be written in a different way by noting that the events $\sbrrr{ T_{m+\eta} - T_m < \tau} = \sbrrr{U\of{T_m+\tau} \ge m+\eta} = \sbrrr{ \Delta U\of{T_m;T_m + \tau} \ge \eta }$ are equivalent.
		Using again Eq.~\eqref{eq:ThetaEst}, but reordering the sums in a different way, we arrive at:
		\begin{align*}
			&
			\CondEA{
				\brrr{ \Sum{m\ne n}{} \xi_m \theta_{m,n} \xi_n }^2
			}{N}
			\\ \le &
			\br{ \frac{2 \lambda^2\tau}{T-\tau} }^2 
			\Sum{m=1}{N-1} \Sum{\eta=1}{N-m} \CondProb{\Delta U\of{T_m,T_m+\tau} \ge \eta}{N}
			\\ =  &
			\br{ \frac{2 \lambda^2\tau}{T-\tau} }^2 
			\Sum{m = 1}{N-1} \CondEA{ \Delta U\of{T_m,T_m+\tau} }{N}
			.
		\end{align*}
		The appearing expectation is of course the average number of jumps in a $\tau$-interval starting with a renewal under the rather involved condition $U\of{T} = N$.
		For CTRW this expression is independent of $m$ and -- as we have shown -- asymptotically independent of the condition.
		Hence the sum evaluates to $N \sEA{U\of{\tau}}$, which is of course $\sLandau{N}$.

		The last representation lacks any reference to the jumping process at all and is valid for arbitrary correlations between the waiting times.
		Whether the TASD is linear in $\tau$ (or in $N$) depends on whether the last expression is negligible compared to $N^2$.
		Roughly speaking this is true for all random walks that do not ``accumulate'' their jumps in one $\tau$-interval.
		This seems to be the case for a plethora of other processes besides CTRW.
		Although we are not able to prove it rigorously, it appears to hold for the random trap model, whence the second plot of FIG.~\ref{fig:Plot}.
		Here the jumps can not accumulate, because there is a minimal time between the jumps that is determined by the shallowest trap.

	\section{Summary and Conclusion}
		We discussed the time averaged squared displacement of a CTRW with power-law waiting time distribution for a given time lag in its single realizations.
		We have shown that this quantity grows linearly in the lag time $\tau$ as well as in the number of steps performed during the measurement, $N$.
		The latter is a consequence of the independence of increments, the former is due to the walker's immobility in between the jumps.
		Non-linear behavior in $\tau$ comes from the accumulation of jumps, that is not strong enough in CTRW to be relevant.

		For given $N$, the time averaged diffusivity's relative fluctuations have been shown to decay with the square-root of the number of steps, i.e. become small when the data acquisition time and therefore $N$ get large. 

		As we made no references to the distribution of the first waiting time, $T_1$, all our results hold as well for aging CTRW.
		Parts of our arguments hold for CTRW with correlated jumps and for CTRW with correlated waiting times.

	\acknowledgments
		This work is supported by Deutsche Forschungsgemeinschaft (DFG) (project SO 307/4-1).

	\bibliographystyle{aipnum4-1}
	\bibliography{Article,Book,Self,NotRead}
\end{document}